\documentclass[twoside,titlepage,final,12pt]{article}
\usepackage{latexsym}
\usepackage{amssymb}

\usepackage{graphicx}
\usepackage[pdftex,dvipsnames,usenames]{color}
\definecolor{IITred}{rgb}{0.5,0.05,0.05}

\usepackage[colorlinks,citecolor=MidnightBlue,urlcolor=Sepia,linkcolor=ForestGreen]{hyperref}

\catcode`\@=11
\def\ps@ppt{\def\@oddhead{\qquad LHC 
Physics Potential \hfil \thepage\qquad}\def\@evenhead{\qquad\thepage \hfil {Chris Quigg} \qquad}
\def\@oddfoot{}\def\@evenfoot{}}    
\catcode`\@=12

\def\urll#1#2{\mbox{\href{#1}{\sf #2}}}

\makeatletter
   \renewcommand{\section}{\@startsection{section}{1}{0mm}
   {\baselineskip}%
   {\baselineskip}{\normalfont\normalsize\centering}}%
   \makeatother

\newcommand{\gev}{\ensuremath{\hbox{ GeV}}}
\newcommand{\tev}{\ensuremath{\hbox{ TeV}}}
\newcommand{\pb}{\ensuremath{\hbox{ pb}}}
\newcommand{\fb}{\ensuremath{\hbox{ fb}}}
\def\phystoday#1#2#3#4{\frenchspacing{\it Phys. Today }{\bf #1}, #2 (\ifcase#3\or January\or 
         February\or March\or April\or May\or June\or July\or August\or 
         September\or October\or November\or December\fi, 19#4)}
\def\slashii#1{\setbox0=\hbox{$#1$}             % set a box for #1
   \dimen0=\wd0                                 % and get its size
   \setbox1=\hbox{\sl/} \dimen1=\wd1            % get size of /
   \ifdim\dimen0>\dimen1                        % #1 is bigger
      \rlap{\hbox to \dimen0{\hfil\sl/\hfil}}   % so center / in box
      #1                                        % and print #1
   \else                                        % / is bigger
      \rlap{\hbox to \dimen1{\hfil$#1$\hfil}}   % so center #1
      \hbox{\sl/}                               % and print /
   \fi}                                         %
\begin{document}
\begin{flushright}
	FERMILAB--FN--0839--T \\ August 24, 2009
\end{flushright}
\vspace*{\stretch{1}}
%\HRule
\begin{center}
	{\Huge  LHC Physics Potential \textit{vs.}~Energy} 
	%\\ {\Huge Tools for Laboratory Directors} 
	\\[5mm]
	{\large Chris Quigg* } \\[5mm]
	Theoretical Physics Department \\
	Fermi National Accelerator Laboratory\\ Batavia, Illinois 60510 USA \\[8mm]
	\parbox{4.in}{Parton luminosities are convenient for estimating how the physics potential of Large Hadron Collider experiments depends on the energy of the proton beams. I present parton luminosities, ratios of parton luminosities, and contours of fixed parton luminosity for $gg$, $u\bar{d}$, and $qq$ interactions over the energy range relevant to the Large Hadron Collider, along with example analyses for specific processes.}	
\end{center}
%\HRule
\vspace*{\stretch{1.5}}
%%%%%%%%%%%%%%%%%%%%%%%%%%%%%%%
%							  %
%	\begin{center}			  %
%							  %
%							  %
%	\vspace*{\stretch{0.5}}	  %
%	\end{center}			  %
%							  %
%%%%%%%%%%%%%%%%%%%%%%%%%%%%%%%
		*\urll{mailto:quigg@fnal.gov}{E-mail:quigg@fnal.gov}
%\twocolumn
\newpage
\setlength{\parindent}{2ex}
\setlength{\parskip}{12pt}
\noindent
The CERN Laboratory \href{http://user.web.cern.ch/user/news/2009/090806-LHC-restart-energy.html}{announced} on August 6 that ``[t]he Large Hadron Collider will run for the first part of the 2009-2010 run at $3.5\tev$ per beam, with the energy rising later in the run.'' Many particle physicists and other interested observers wish to understand how operating the LHC  below the 14-TeV design energy will affect the initial physics program and the early discovery potential of the ATLAS and CMS experiments. The first objective of the run is to commission and ensure stable operation of the accelerator complex and the experiments. For the experiments, an essential task is to ``rediscover'' the standard model of particle physics, and to use familiar physics objects such as $W^\pm$, $Z^0$, $J\!/\!\psi$, $\Upsilon$, jets, $b$-hadrons, and top-quark pairs to tune detector performance. Within reasonable limits, energy and luminosity will not be critical parameters in the first weeks and months, and it is natural to expect that they will be chosen in light of accelerator performance.

As commissioning leads into a first physics run, two questions assume great interest:
\vspace*{-18pt}
\begin{itemize}
\item How is the physics potential compromised by running below $14\tev$?
%\vspace*{-12pt}
\item At what point will the LHC begin to explore virgin territory and surpass the discovery reach of the Tevatron experiments CDF and D0?
\end{itemize}
\vspace*{-18pt}
Detailed simulations of signals and backgrounds are of unquestioned value for in-depth consideration of the physics possibilities. However, much can be learned about the general issues of energy, luminosity, and the relative merits of proton-proton and proton-antiproton collisions by comparing the luminosities of parton-parton collisions as a function of $\sqrt{\hat{s}}$, the c.m. energy of the colliding partons~\cite{Eichten:1984eu}. [A high-energy proton is, in essence, a broadband unseparated beam of quarks, antiquarks, and gluons.] By contemplating the parton luminosities in light of existing theoretical and experimental knowledge, physicists should be able to anticipate and critically examine the broad results of Monte Carlo studies. The more prior knowledge a user brings to the parton luminosities, the more useful insights they can reveal.

Taking into account the $1/\hat{s}$ behavior of the hard-scattering processes that define much of the physics motivation for a multi-TeV hadron collider, the \textit{parton luminosity} 
\begin{equation}
\frac{\tau}{\hat{s}}\frac{d\mathcal{L}_{ij}}{d\tau}  \equiv  \frac{\tau/\hat{s}}{1 + \delta_{ij}}\int_\tau^1 \!\!dx[f_i^{(a)}(x) f^{(b)}_j(\tau/x)  + f_j^{(a)}(x)f_i^{(b)}(\tau/x)]/x, 
 \label{eq:lumdef}
 \end{equation}
 which has dimensions of a cross section, is a convenient measure of the reach of a collider of given energy and hadron-hadron luminosity. Here $f_i^{(a)}(x)$ is the number distribution of partons of species $i$ carrying momentum fraction $x$ of hadron $a$. For hadrons colliding with c.m. energy $\sqrt{s}$, the scaling variable $\tau$ is given by
 \begin{equation} 
 \tau  = \hat{s}/s .
 \label{eq:taudef}
 \end{equation}
 
The  cross section for the hadronic reaction
 \begin{equation}
 a + b \to \alpha + \hbox{anything}
 \label{eq:abcX}
 \end{equation}
 is given by 
 \begin{equation}
\sigma(s) = \sum_{\{ij\}}\int_{\tau_0}^1 \frac{d\tau}{\tau}\cdot{\frac{\tau}{\hat{s}}\frac{d\mathcal{L}_{ij}}{d\tau}}\cdot\left[\hat{s}\hat{\sigma}_{ij \to \alpha}(\hat{s})\right] ,
\label{eq:cross}
\end{equation}
 where $\hat{\sigma}_{ij \to \alpha}$ is the operative parton-level cross section. The (dimensionless) factor in square brackets is approximately determined by couplings. Many explicit (leading-order) forms of $\hat{\sigma}$ are given in Refs.~\cite{Eichten:1984eu}. The logarithmic integral typically gives a factor of order unity.

 If event rates for signal and backgrounds are known---by calculation or by measurement---for some point $(\sqrt{s},\sqrt{\hat{s}})$, the parton luminosities can be used to estimate the rates at other points, at an accuracy satisfactory for orientation. Because leading-order parton distributions have a simple intuitive interpretation, I have chosen the CTEQ6L1 leading-order parton distributions~\cite{Pumplin:2002vw} for the calculations presented in this note.

\begin{figure}[tb]
%\psfrag{w}[t]{$\sqrt{\hat{s}}$}
\centerline{\includegraphics[height=0.7\textheight]{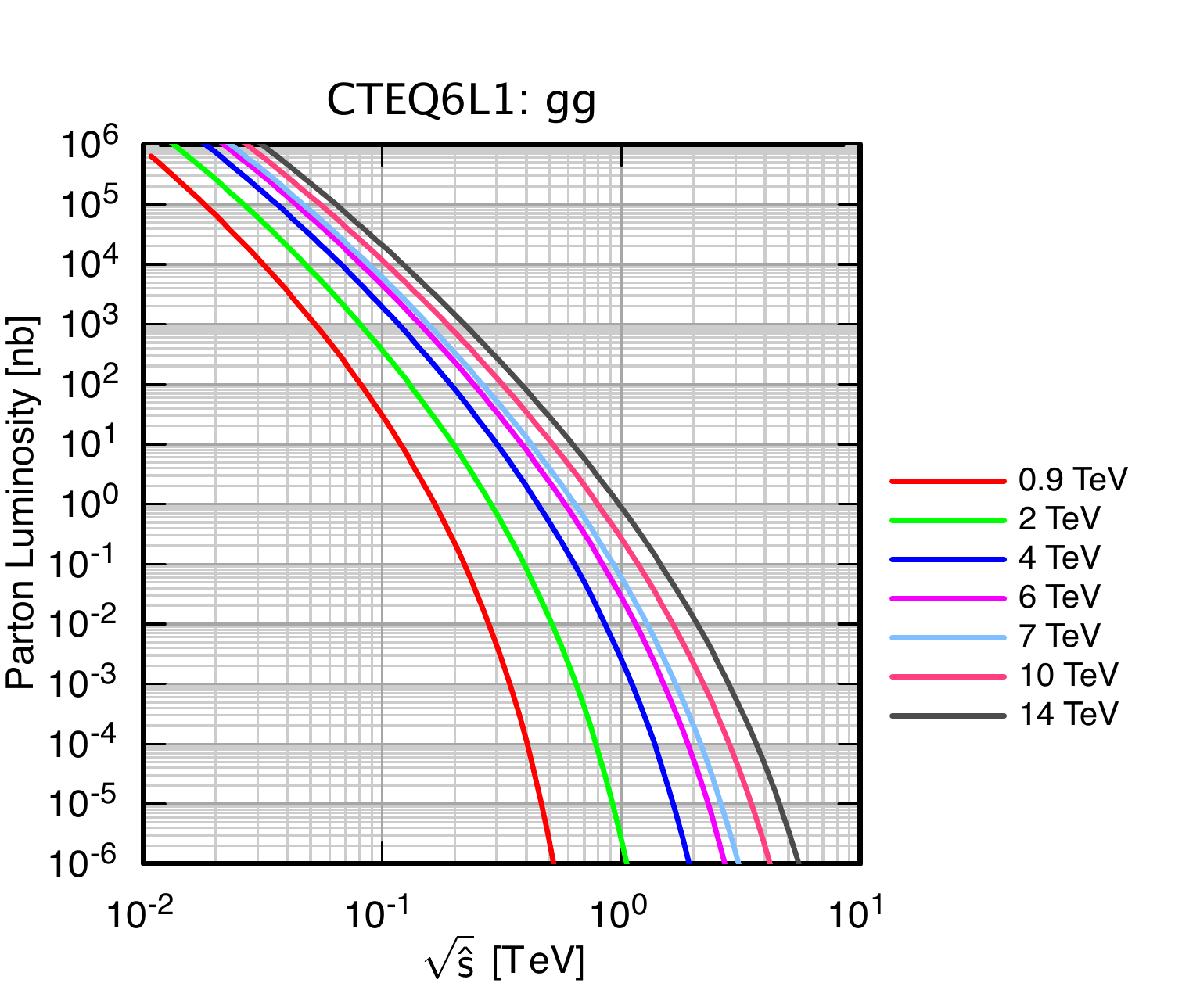}}
\caption{Parton luminosity $(\tau/\hat{s})d\mathcal{L}/d\tau$ for $gg$ interactions.}
\label{fig:gg2}
\end{figure}

\begin{figure}[tb]
\centerline{\includegraphics[height=0.7\textheight]{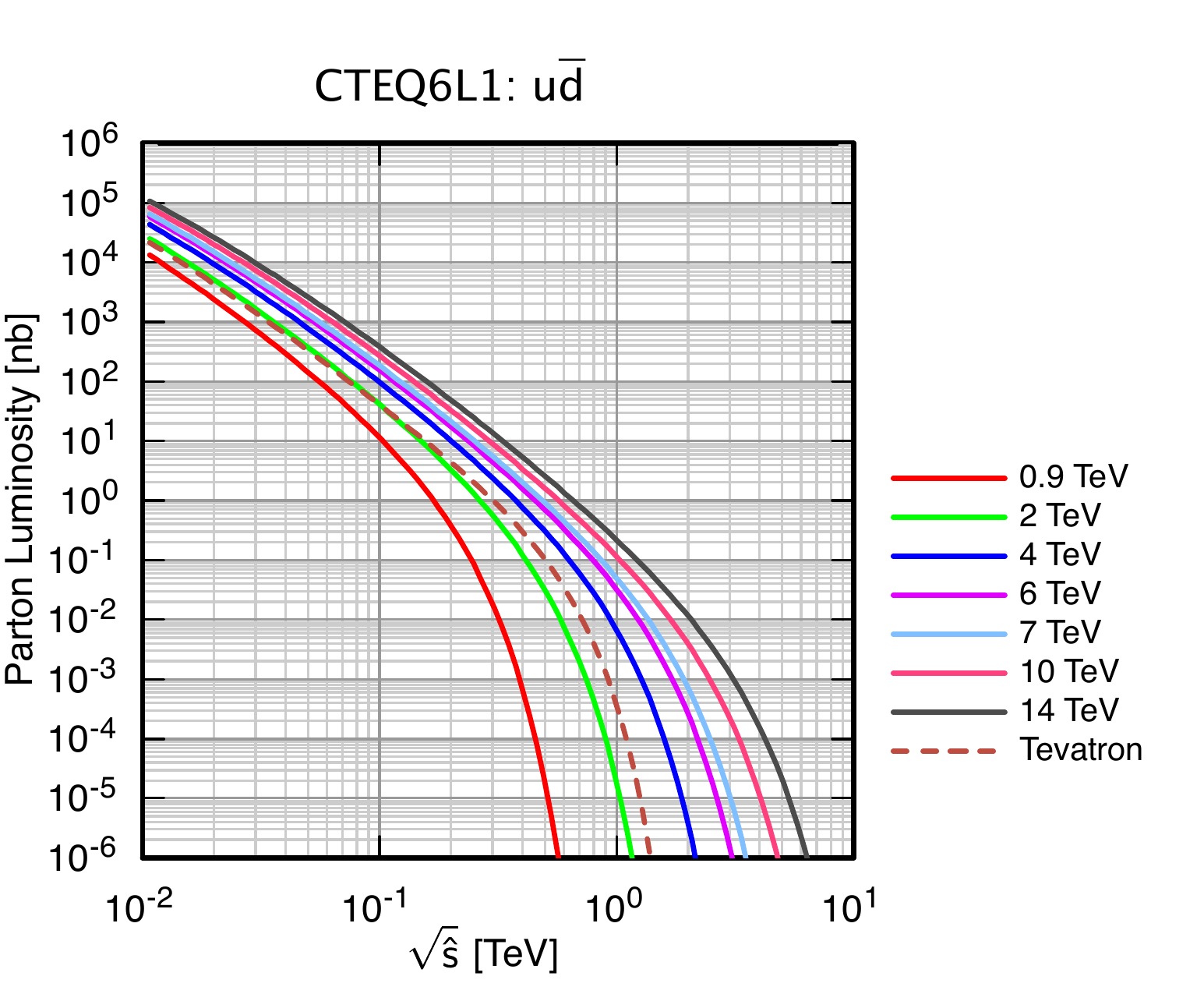}}
\caption{Parton luminosity $(\tau/\hat{s})d\mathcal{L}/d\tau$ for $u\bar{d}$ interactions.}
\label{fig:udbar2}
\end{figure}

\begin{figure}[tb]
\centerline{\includegraphics[height=0.7\textheight]{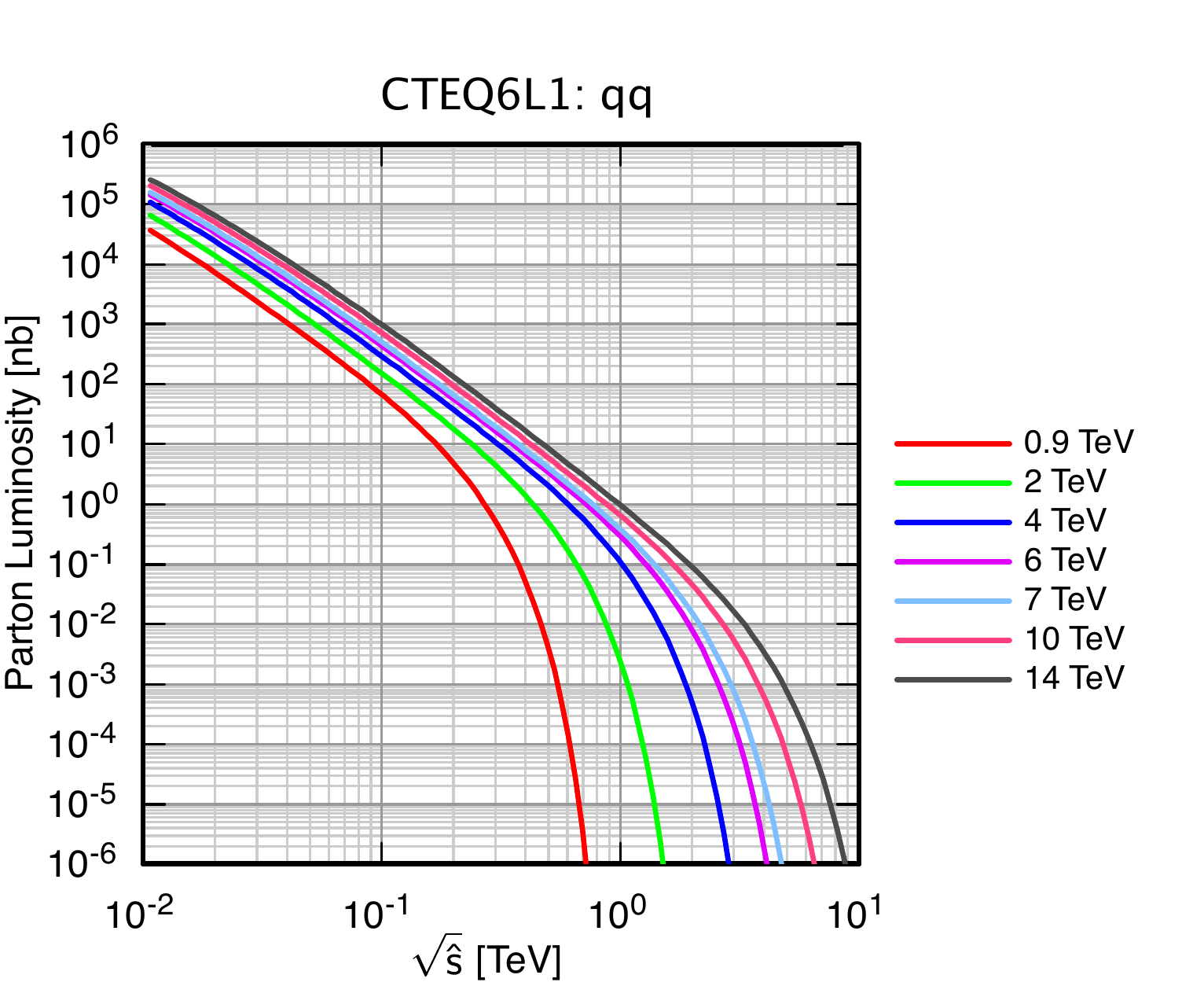}}
\caption{Parton luminosity $(\tau/\hat{s})d\mathcal{L}/d\tau$ for $qq$ interactions.}
\label{fig:qq2}
\end{figure}

Three examples suffice to survey the important large-cross section processes most relevant to early running: gluon-gluon interactions, $u\bar{d}$ interactions, and interactions among generic light quarks. The parton luminosities for gluon-gluon interactions are given in Figure~\ref{fig:gg2}. These are identical for $pp$ and $\bar{p}p$ collisions. The parton luminosities for $u\bar{d}$ interactions are plotted in Figure~\ref{fig:udbar2}. In $pp$ collisions, $u\bar{d}$ is a valence--sea combination; in $\bar{p}p$ collisions, it is valence--valence. The difference is reflected in the excess of the Tevatron luminosities in Figure~\ref{fig:udbar2} over the proton-proton luminosities at $\sqrt{s}= 2\tev$.
The parton luminosities for light-quark--light-quark interactions in $pp$ collisions are displayed in Figure~\ref{fig:qq2}, as examples of valence--valence interactions leading to final states such as two jets. What is plotted here is the combination
\begin{equation}
(u + d)^{(1)}\otimes (u + d)^{(2)} .
\label{eq:qqdef}
\end{equation}
For $\bar{p}p$ collisions at the Tevatron, interpret the 2-TeV curve as
\begin{equation}
(u + d)^{(p)}\otimes (\bar{u} + \bar{d})^{(\bar{p})} ;
\label{eq:qqdef}
\end{equation}
these valence--valence interactions are the main source of high-transverse-momentum jets at the Tevatron.

\textit{Ratios} of parton luminosities are especially useful for addressing the two questions. Let us consider each of the example cases in turn.

\begin{figure}[tb]
\centerline{\includegraphics[height=0.7\textheight]{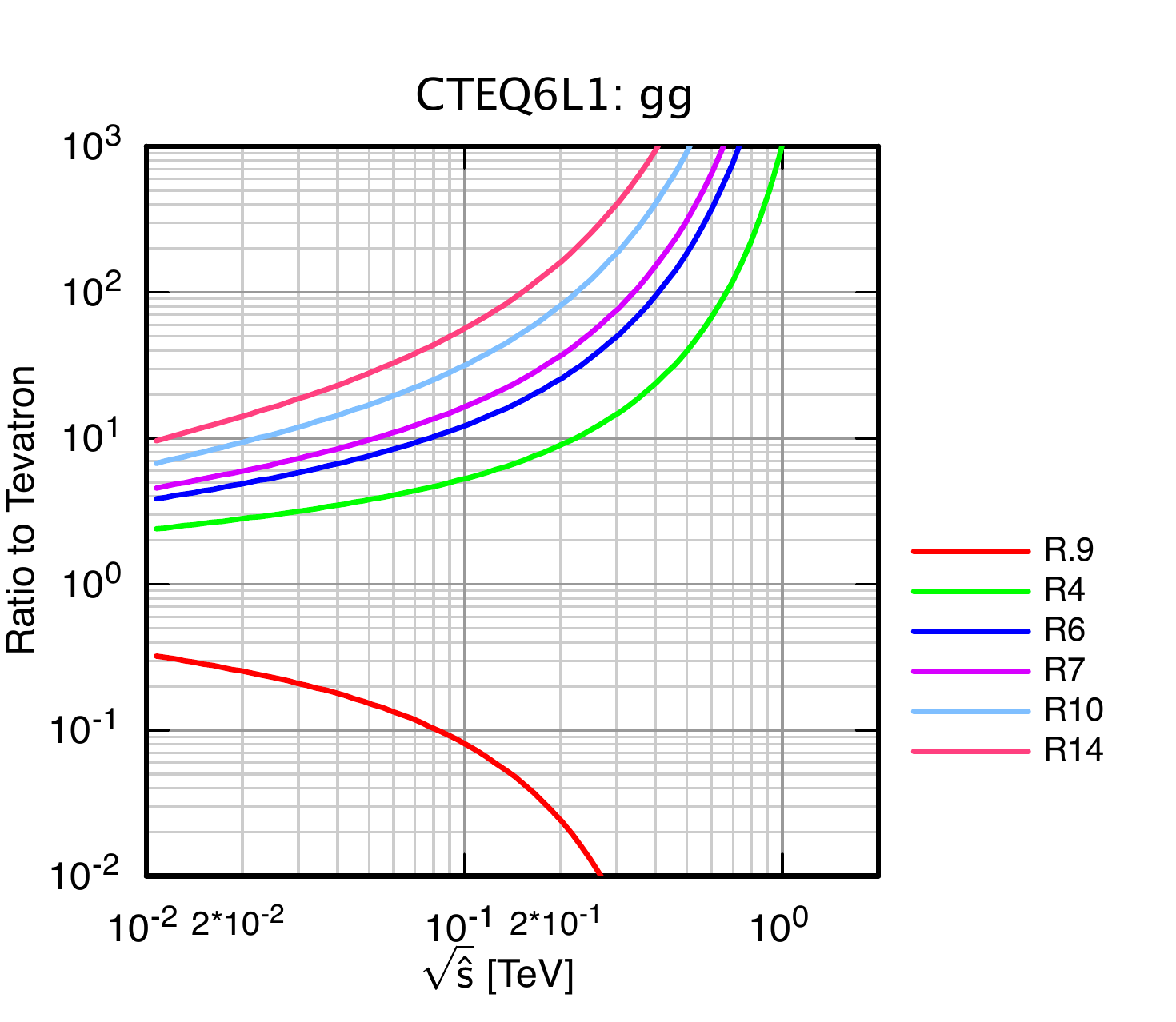}}
\caption{Comparison of parton luminosity for $gg$ interactions at specified energies with luminosity at $2\tev$.}
\label{fig:ggrat2}
\end{figure}

\begin{figure}[tb]
\centerline{\includegraphics[height=0.7\textheight]{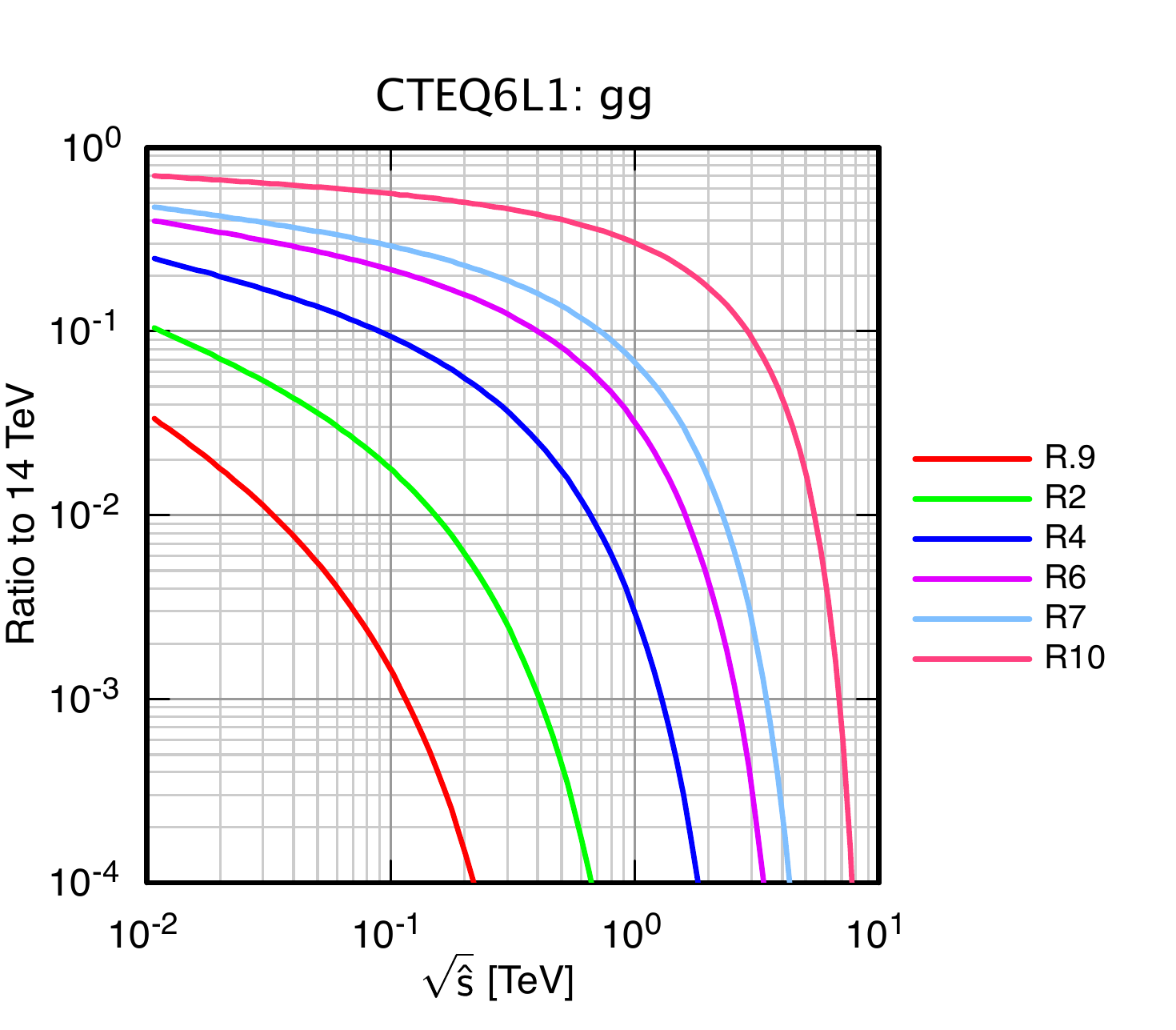}}
\caption{Comparison of parton luminosity for $gg$ interactions at specified energies with luminosity at $14\tev$.}
\label{fig:ggrat14}
\end{figure}

Ratios of parton luminosities for gluon-gluon interactions in $p^\pm p$ collisions at specified energies to the $gg$ luminosity at the Tevatron are shown in Figure~\ref{fig:ggrat2}; ratios to the LHC at design energy in Figure~\ref{fig:ggrat14}. At $\sqrt{\hat{s}} \approx 0.4\tev$, characteristic of $t\bar{t}$ pair production, Figure~\ref{fig:ggrat2} shows that the $gg$ luminosity rises by three orders of magnitude from the Tevatron at $2\tev$ to the LHC at $14\tev$. This rise is the source of the computed increase in the $gg \to t\bar{t}$ cross section from Tevatron to LHC, and is the basis for the (oversimplified) slogan,  ``The Tevatron is a quark collider, the LHC is a gluon collider." Figure~\ref{fig:ggrat14} shows that the $gg \to t\bar{t}$ yield drops by a bit more than a factor of 6 between $14\tev$ and $7\tev$. To first approximation, accumulating a $t\bar{t}$ sample of specified size at $\sqrt{s}=7\tev$ will require about $6\times$ the integrated luminosity that would have been needed at $\sqrt{s}=14\tev$, although acceptance cuts should have less effect at the lower energy. At $\sqrt{s} = 10\tev$, the $gg \to t\bar{t}$ rate is a factor of $2.3$ smaller than at design energy.

The dominant mechanism for light Higgs-boson production at both the Tevatron and the LHC is $gg \to \hbox{top-quark loop} \to H$, so the rates are controlled by the $gg$ luminosity. For $M_H \approx 120\gev$, the $gg$ luminosity is approximately $(20, 38, 70)\times$ larger at $\sqrt{s} = (7, 10, 14)\tev$ than at the Tevatron. LHC experiments are likely to rely on the rare $\gamma\gamma$ decay of a light Higgs boson, for which the rates in early low-luminosity running are too small to matter. At somewhat higher Higgs-boson masses, the situation could be more promising for early running. For $M_H = 175\gev$, a mass at which $H \to ZZ$ becomes a significant decay mode, the $gg$ luminosity is roughly $(30, 65, 130)\times$ larger at $\sqrt{s} = (7, 10, 14)\tev$ than at the Tevatron. The potential Tevatron sensitivity for $gg \to H \to ZZ$, based on the current integrated luminosity of $6\fb^{-1}$ would be matched at the LHC by integrated luminosities of $(200, 90, 45)\pb^{-1}$ at $\sqrt{s} = (7, 10, 14)\tev$. Note that these levels do not correspond to the thresholds needed for discovery (although those could be worked out, given a discovery criterion), but to the point at which the LHC would begin to break new ground, compared to the Tevatron sample now in hand.

\begin{figure}[tb]
\centerline{\includegraphics[height=0.7\textheight]{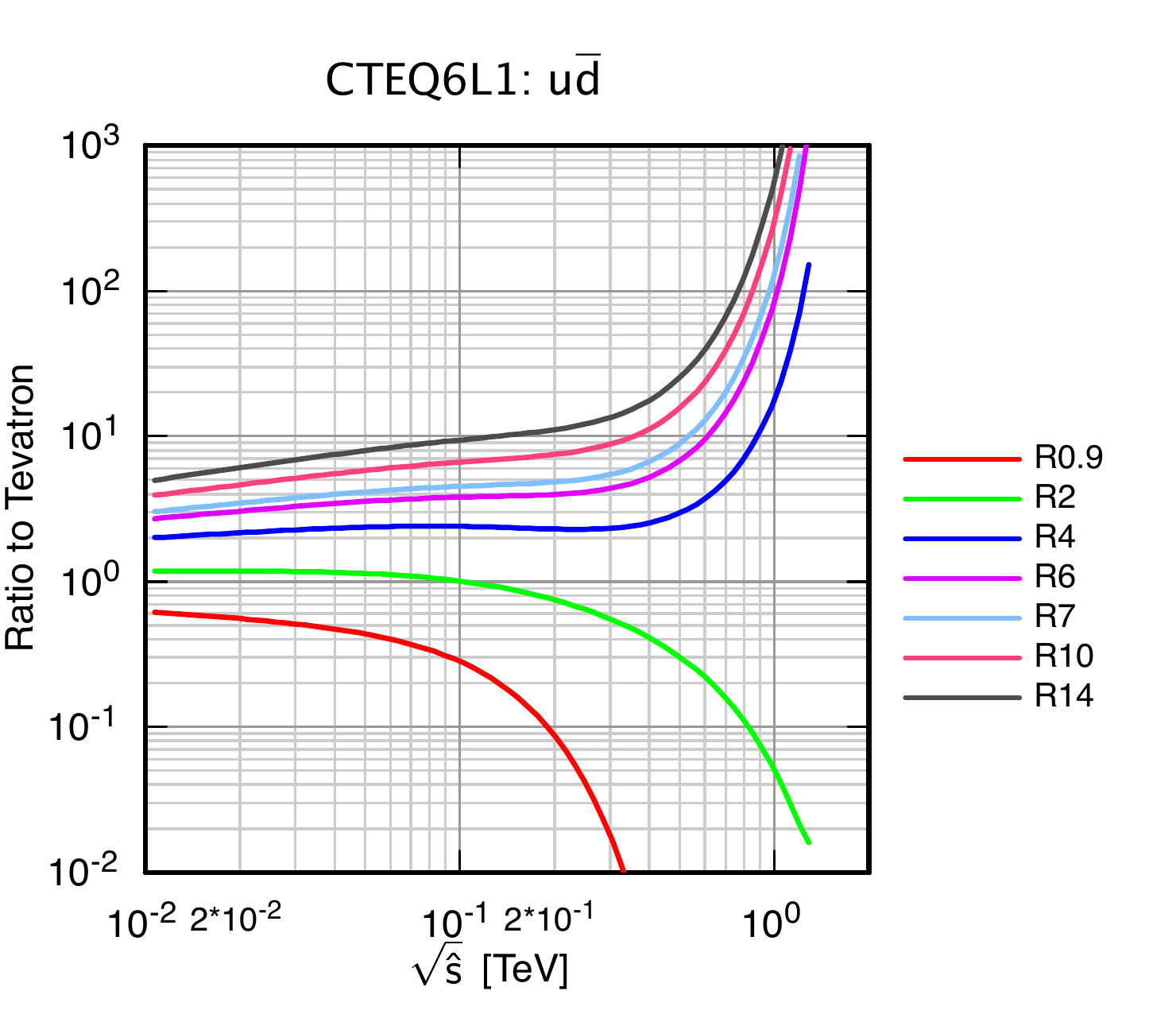}}
\caption{Comparison of parton luminosity for $u\bar{d}$ interactions in $pp$ collisions at specified energies with luminosity in $\bar{p}p$ collisions at $2\tev$.}
\label{fig:udbarrat2}
\end{figure}

\begin{figure}[tb]
\centerline{\includegraphics[height=0.7\textheight]{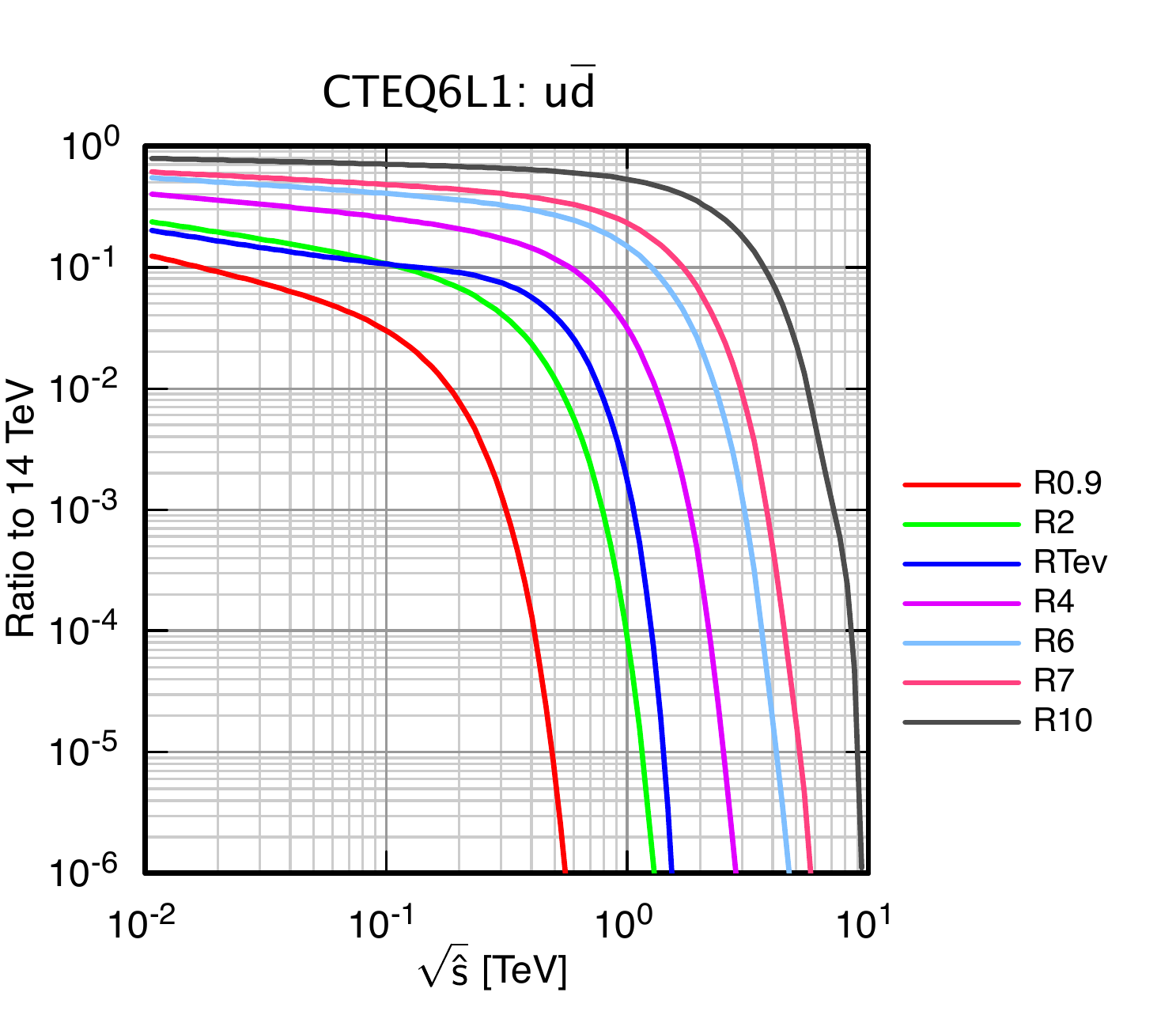}}
\caption{Comparison of parton luminosity for $u\bar{d}$ interactions at specified energies with luminosity at $14\tev$.}
\label{fig:udbarrat14}
\end{figure}

Ratios of parton luminosities for $u\bar{d}$ interactions in $pp$ collisions at specified energies to the $u\bar{d}$ luminosity in $\bar{p}p$ collisions at the Tevatron are plotted in Figure~\ref{fig:udbarrat2}. Ratios to the LHC at design energy are shown in Figure~\ref{fig:udbarrat14}. These ratios of luminosities apply directly to the production of $W$ bosons and to the search for new $W^\prime$ bosons. They are also indicative of the behavior of $u\bar{u}$ and $d\bar{d}$ luminosities, which enter the production of $Z$, $Z^\prime$, and $W^+W^-$ or $ZZ$ pairs that are backgrounds to Higgs-boson searches for $M_H \gtrsim 140\gev$. For the case of $W^+$ production, Figure~\ref{fig:udbarrat2} shows that the rates will be higher by factors of $(4.4, 6.4, 9)$ at $\sqrt{s} = (7, 10, 14)\tev$, compared to the Tevatron rate. At an invariant mass of $175\gev$, the curves in Figure~\ref{fig:udbarrat2} show that the rate for the background processes $q\bar{q} \to VV$ ($V = W,Z$) grows less rapidly than the rate for the signal process $gg \to H$ discussed in the previous paragraph. The enhancements over the Tevatron are by factors of $(4.8,7.3,10.7)$ at $\sqrt{s} = (7, 10, 14)\tev$. 

For a $W^\prime$ search at $M_{W^\prime} = 0.8\tev$, the production rates are larger by factors of $(30, 70, 120)$ at $\sqrt{s} = (7, 10, 14)\tev$, so the current Tevatron sensitivity would be matched at integrated luminosities of approximately $(200, 90, 45)\pb^{-1}$, before taking into account relative detector acceptances. At still higher masses, the penalty for LHC running below design energy is correspondingly greater. At $M_{W^\prime} = 2\tev$, the rates are diminished by factors of approximately $3$ and $17$ at $\sqrt{s}=(10,7)\tev$. For high-mass searches, one must refer back to the parton luminosities themselves (Figure~\ref{fig:udbar2}) to check whether the absolute rates give adequate sensitivity.

The $q\bar{q}$ contribution to $t\bar{t}$ production will also track the ratios of $u\bar{d}$ luminosities. In the range of interest, $\sqrt{\hat{s}} \approx 0.4\tev$, the rate is enhanced over the Tevatron ($\bar{p}p$ collisions!) rate by factors of roughly $(7, 11, 18)$ at $\sqrt{s} = (7, 10, 14)\tev$, far smaller than the enhancements we noted above for the $gg \to t\bar{t}$ rates. The behavior of the $u\bar{d}$ parton luminosities also determines how rates for the pair production of new colored particles (e.g., superpartners) scale under different operating conditions.

\begin{figure}[tb]
\centerline{\includegraphics[height=0.7\textheight]{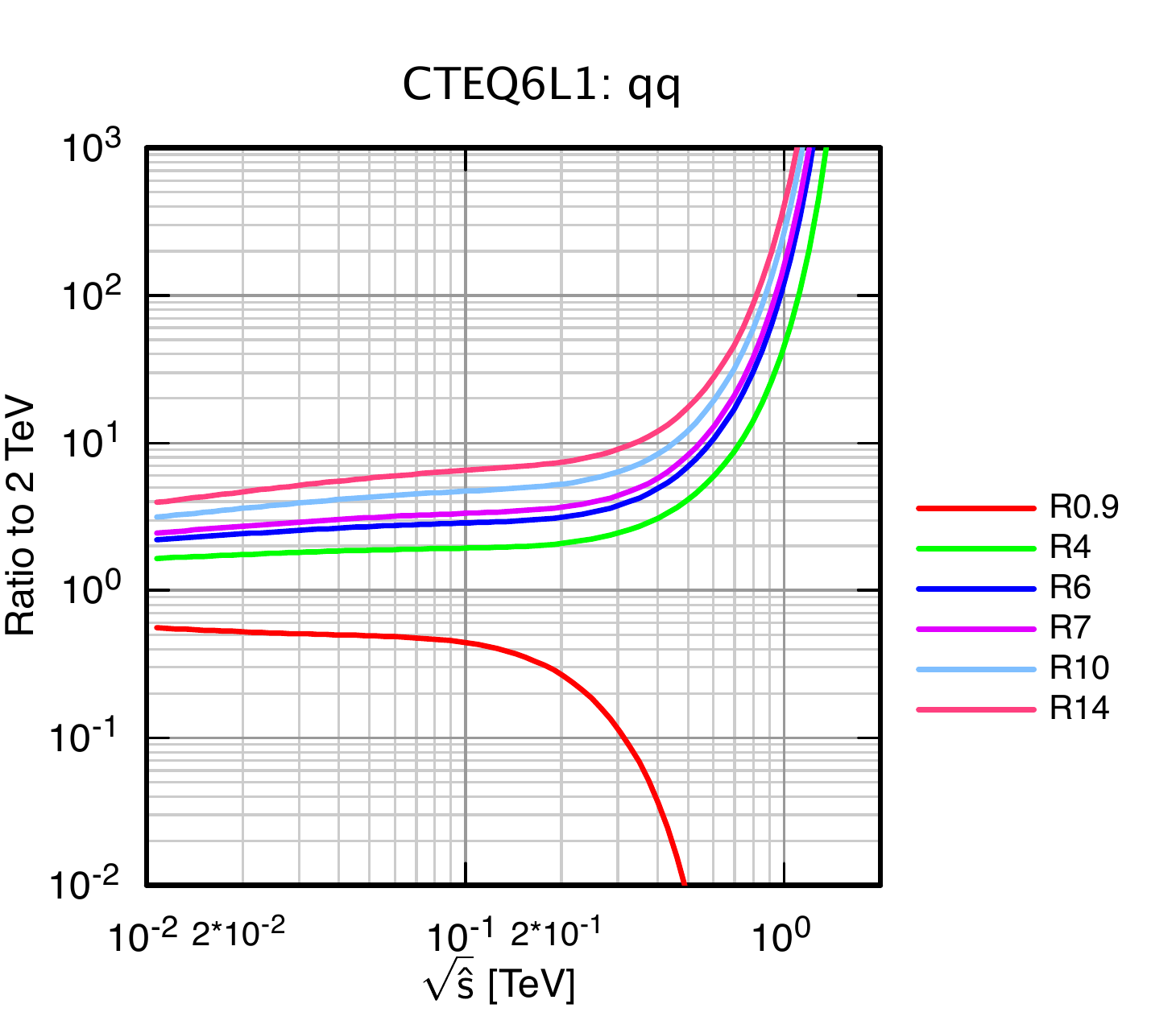}}
\caption{Comparison of parton luminosity for $qq$ interactions at specified energies with luminosity at $2\tev$.}
\label{fig:qqrat2}
\end{figure}

\begin{figure}[tb]
\centerline{\includegraphics[height=0.7\textheight]{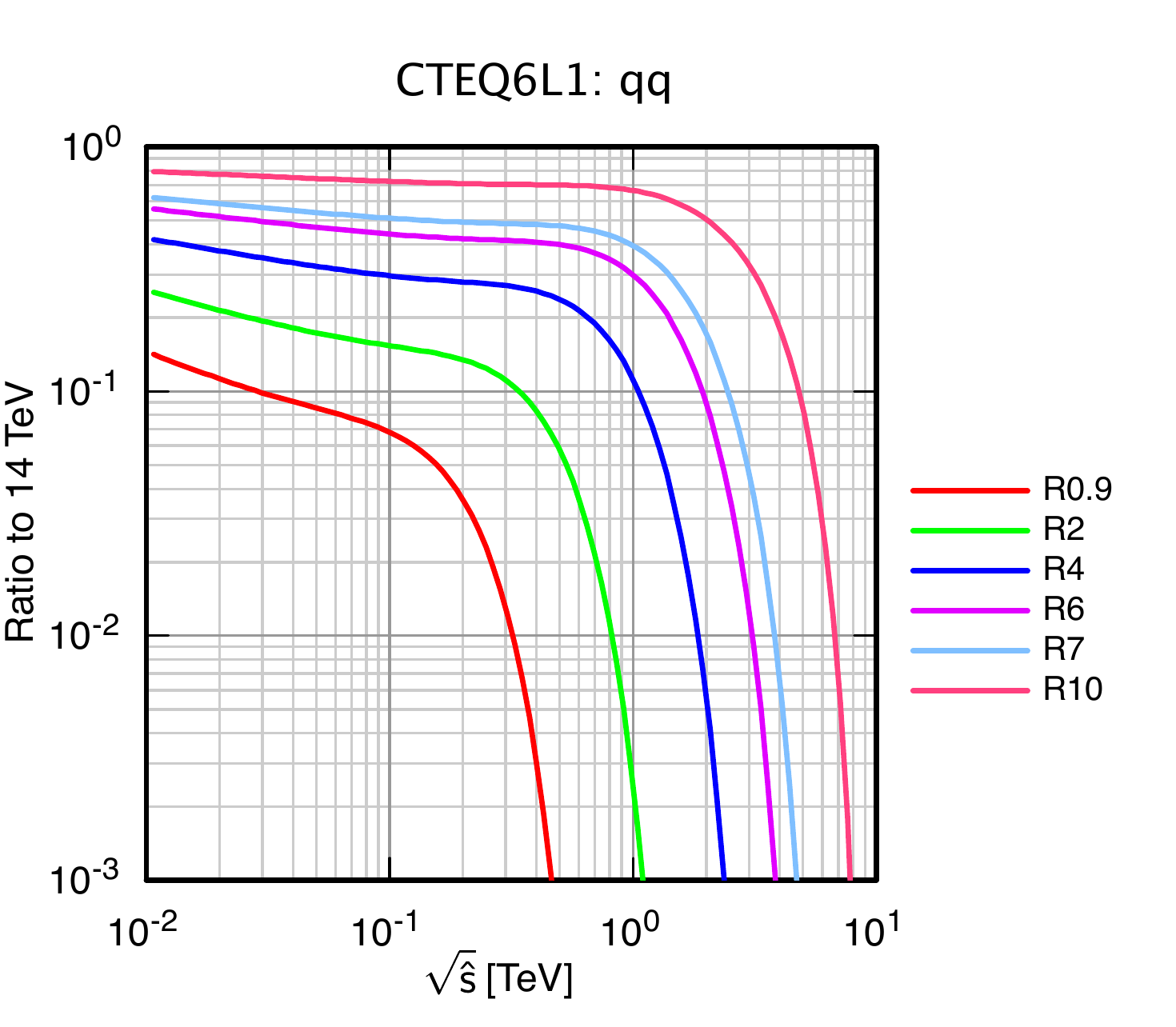}}
\caption{Comparison of parton luminosity for $qq$ interactions at specified energies with luminosity at $14\tev$.}
\label{fig:qqrat14}
\end{figure}

Ratios of parton luminosities for generic light-quark--light-quark interactions in $pp$ collisions at specified energies to the corresponding light-quark--light-antiquark interactions in $\bar{p}p$ collisions at the Tevatron are displayed in Figure~\ref{fig:qqrat2}. Ratios of the same luminosities  to the LHC at design energy appear in Figure~\ref{fig:qqrat14}. Such valence--valence interactions govern the production of hadron jets at very large values of $p_\perp$. Figure~\ref{fig:qqrat2} reveals that for jet production at $\sqrt{\hat{s}} \approx 1\tev$, $qq \to \hbox{two jets}$ will be enhanced at the LHC by factors of $(135, 220, 340)$ over the rate for $q\bar{q} \to \hbox{two jets}$ at the Tevatron. At higher scales, the relevant question is how much rates are diminished in running below $\sqrt{s} = 14\tev$. For $\sqrt{\hat{s}} = 2\tev$, the rates at $\sqrt{s} = (7, 10)\tev$ are $(0.53, 0.17)\times$ those at design energy. The corresponding multipliers at $\sqrt{\hat{s}} = 4\tev$ are $(0.18, 0.008)$.

\begin{figure}[tb]
\centerline{\includegraphics[height=0.7\textheight]{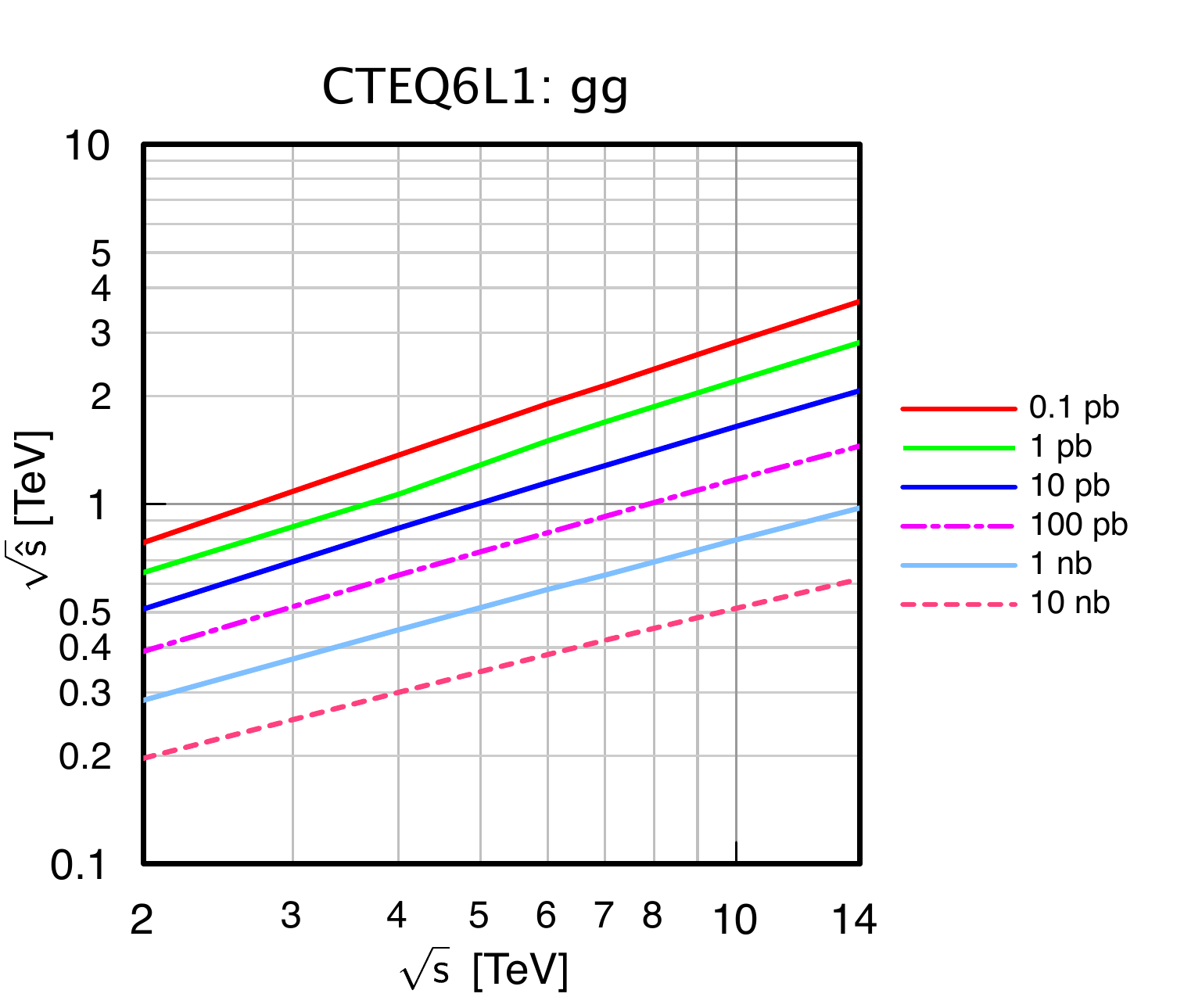}}
\caption{Contours of parton luminosity for $gg$ interactions in $p^\pm p$ collisions.}
\label{fig:ggcontours}
\end{figure}

\begin{figure}[tb]
\centerline{\includegraphics[height=0.7\textheight]{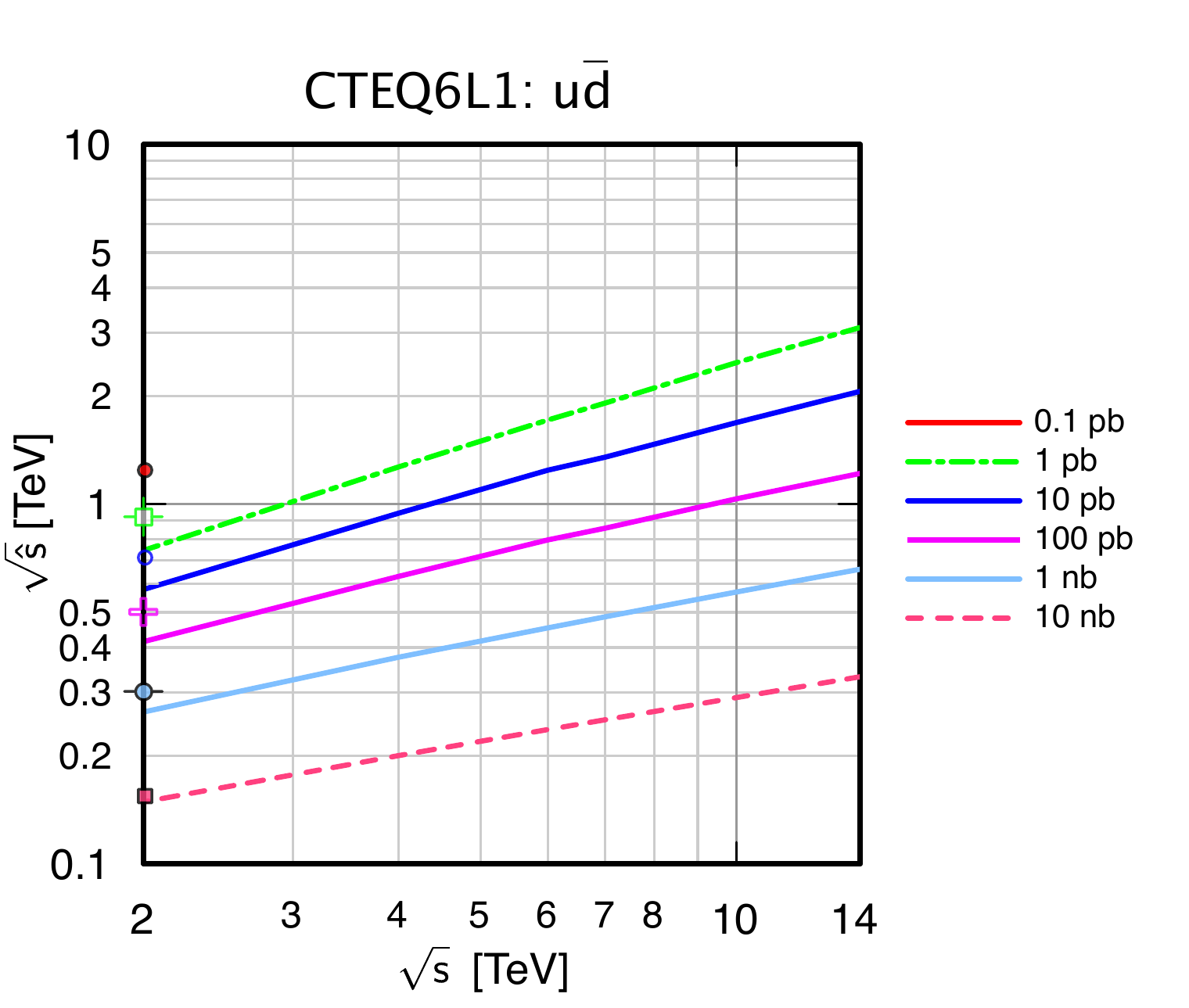}}
\caption{Contours of parton luminosity for $u\bar{d}$ interactions in $pp$ collisions. Values of $w$ corresponding to the stated values for $\bar{p}p$ collisions at the Tevatron are shown as points at $E = 2\tev$.}
\label{fig:udbarcontours}
\end{figure}

\begin{figure}[tb]
\centerline{\includegraphics[height=0.7\textheight]{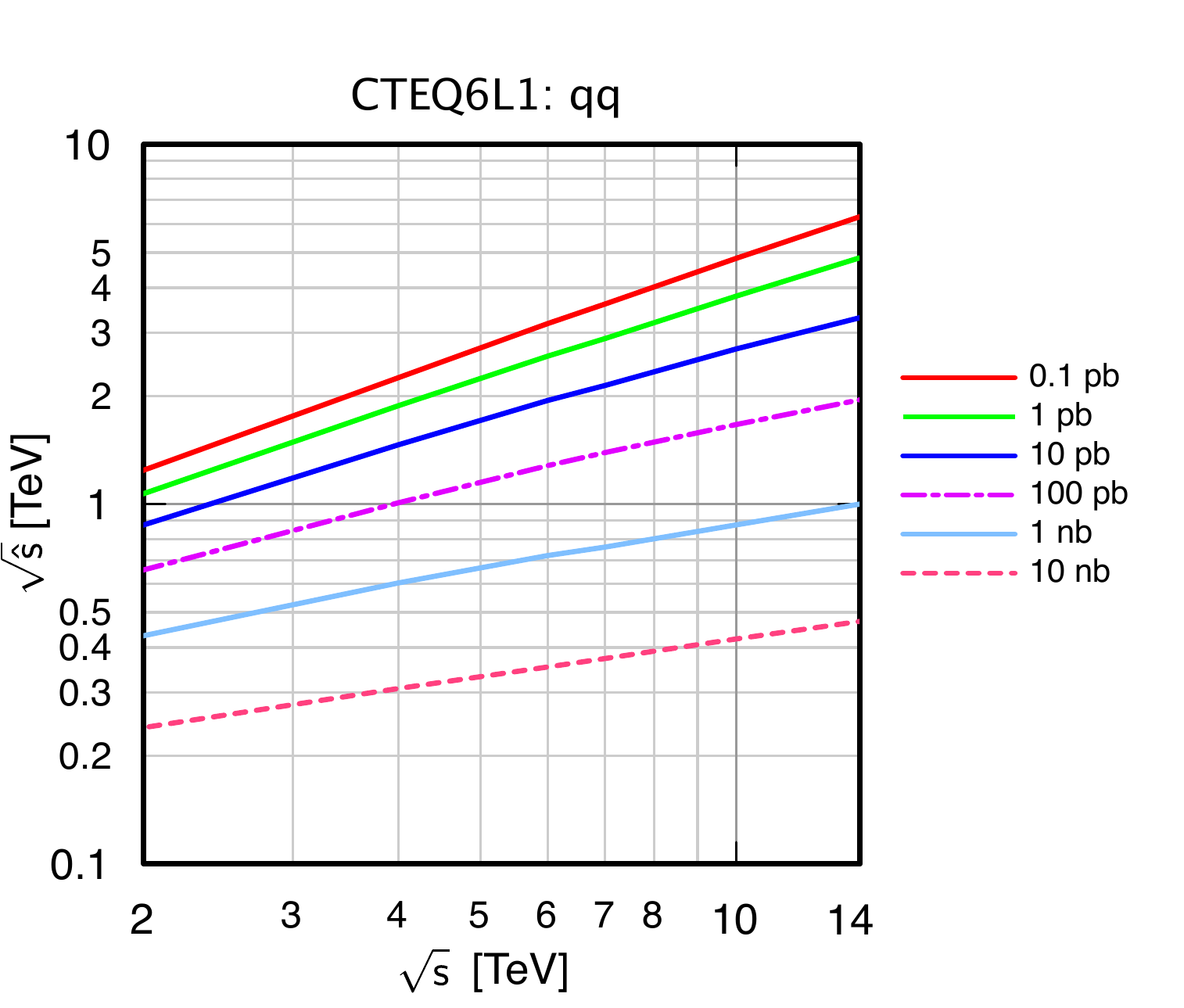}}
\caption{Contours of parton luminosity for $qq$ interactions in $pp$ collisions or $q\bar{q}$ interactions in $\bar{p}p$ collisions.}
\label{fig:qqcontours}
\end{figure}

Contour plots showing at each proton-(anti)proton energy $\sqrt{s}$ the parton-parton energy $\sqrt{\hat{s}}$ that corresponds to a particular value of parton luminosity $(\tau/\hat{s})d\mathcal{L}/d\tau$ provide another tool for judging the effects of changes in beam energy or proton-proton luminosity. Plots for the three examples considered in this note are given in Figures~\ref{fig:ggcontours}, \ref{fig:udbarcontours}, and \ref{fig:qqcontours}, for proton-proton energies between $2$ and $14\tev$. The advantage of 2-TeV $\bar{p}p$ collisions over 2-TeV $pp$ collisions for $u\bar{d}$ interactions is indicated by the Tevatron points in Figure~\ref{fig:udbarcontours}.
The contour plots summarize a great deal of information, and will reward detailed study. As for the parton luminosity and ratio plots, study of the contour plots will be particularly informative to the user who brings a thorough understanding of signals and backgrounds at one or more beam energies.

{\footnotesize
%\vspace*{12pt}
\noindent I thank Estia Eichten and Ken Lane for valuable advice, and Bogdan Dobrescu, Keith Ellis,  and Paddy Fox for comments on the manuscript. Fermilab is operated by the Fermi Research Alliance under contract no.\  DE-AC02-07CH11359 with the
U.S.\ Department of Energy.

\bibliographystyle{utphysrevA}

\bibliography{Luminosities}

}
 

\end{document}